\documentclass[iop,revtex4]{emulateapj}
\usepackage{amstext,amsmath,natbib}
\usepackage{epstopdf}
\usepackage{multirow,lscape}
\usepackage{graphicx, graphics}
\usepackage{color}

\shorttitle{\ion{O}{7} absorption lines in x-ray targets}
\shortauthors{Luo et al.}

\begin{document}

\title{\sl{XMM}-Newton Survey of Local \ion{O}{7} Absorption Lines in the Spectra of Galactic X-ray sources}

\author{Yang~Luo\altaffilmark{1,2,3}, Taotao~Fang\altaffilmark{1}, Renyi~Ma\altaffilmark{1}}

\altaffiltext{1}{Department of Astronomy and Institute of Theoretical Physics and
  Astrophysics, Xiamen University, Xiamen, Fujian 361005, China; fangt@xmu.edu.cn} 
\altaffiltext{2}{Theoretical Astrophysics, Department of Earth \& Space Science, Osaka University,
     1-1 Machikaneyama, Toyonaka, Osaka 560-0043, Japan}
\altaffiltext{3}{Department of Physics \& Astronomy, University of Kentucky, Lexington, KY 40506-0055, USA}
\begin{abstract}

The detection of highly ionized metal absorption lines in the X-ray spectra of the Galactic X-ray binaries (XRBs) implies the distribution of hot gas along the sightline toward the background sources. However, the origin of this hot gas is still unclear: it can arise in the hot interstellar medium (ISM), or is intrinsic to the XRBs. In this paper, we present an {\sl XMM}-Newton survey of the \ion{O}{7} absorption lines in the spectra of Galactic XRBs. A total of 33 XRBs were selected, with 29 low mass XRBs and 4 high mass XRBs. At more than 3$\sigma$ threshold, \ion{O}{7} absorption line was detected in 16 targets, among which 4 were newly discovered in this work. The average line equivalent width (EW) is centered around $\sim$ 20 m\AA. Additionally we do not find strong correlations between the \ion{O}{7} EWs and the Galactic neutral absorption N$_{\rm H}$, the Galactic coordinates or the distance of background targets. Such non-correlation may suggest the contamination of the circumstellar material, or the lack of constrains on the line Doppler-b parameter. We also find regardless the direction of the XRBs, the \ion{O}{7} absorption lines were always detected when the flux of the background XRBs reach a certain level, suggesting a uniform distribution of this hot gas. We estimate a ratio of 0.004 --- 0.4 between hot and neutral phase of the interstellar medium. This is the second paper in the series following \citet{fang2015}, in which we focused on the local \ion{O}{7} absorption lines detected in the background AGN spectra. Detailed modeling of the hot ISM distribution will be investigated in a future paper.

\end{abstract}

\keywords{galaxies: ISM ---  X-rays: ISM ---  ISM: general} 

\section{Introduction}

The presence of hot interstellar medium (ISM) has been predicted for a long time \citep[for a review, see][]{spitzer1956,ferriere2001,putman2012}. This hot gas ($\sim$10$^6$K) could be a result of stellar feedback (supernova explosions or stellar wind from massive stars) or shock-heated intergalactic medium (IGM) accreting to the Galactic center, therefore study of the hot ISM could have a profound implication on the theory of galaxy formation and evolution. 

Current understanding of the Galactic hot gas initially comes from various broad-band observations of the diffuse soft X-ray background. At the soft X-ray band, other than the extragalactic component, the solar wind charge exchange and the local hot bubble component, a contribution of the background is expected from a large-scale hot plasma present in the Galactic halo \citep[e.g.,][]{Lallement2004,Cappelluti2012,henley2013,Galeazzi2014}. Insights on the properties of the hot ISM also come from the study of z=0, metal absorption lines in extragalactic active galactic nuclei (AGN). The detection of z=0, highly ionized absorption lines (e.g, \ion{Ne}{9}, \ion{O}{7} and \ion{O}{8}) toward bright extragalactic sources \citep[e.g.,][]{nicastro2002,fang2003,rasmussen2003} and their Galactic origin \citep{fang2006,bregman2007} also suggested the existence of the hot ISM. 

However, current data cannot give a precise description of the spatial distribution of the hot gas. Evidence from the disk-morphology of the X-ray emission of nearby galaxies is suggestive of the gas to be confined in the disk \citep{tullmann2006,li2013}. However, semi-analytic calculations and numerical simulations for disk galaxy formation predict the existence of extended hot gaseous halos around massive spiral galaxies due to the accretion of the intergalactic medium \citep[e.g.,][]{White1991,Mo2002,maller2004,Fukugita2006,Crain2010,Ntormousi2010,Marinacci2011,Stinson2012,Feldmann2013,fang2013,sokoowska2016}. The hot ISM may extend out to at least the Magellanic System, as suggested by the pressure confinement of high-velocity cloud or satellite galaxies stripping stream \citep{sembach2003,Collins2005,Grcevich2009,lehner2012,Gatto2013}. Detections of diffuse X-ray halos around massive normal spiral galaxies also suggested the existence of hot, X-ray emitting gas beyond the disk \citep{anderson2011,dai2012,bogdan2012,bogdan2013,hodges-kluck2013,anderson2016}. 

In the past few years, it has become well established that a large amount of hot gas is present along the sightline toward background XRBs. Located in the Galactic disk, the XRBs provide a potential probe of the ISM in the disk. The detection of X-ray absorption lines of highly ionized species (e.g, \ion{Ne}{9}, \ion{O}{7} and \ion{O}{8}) toward the XRBs leads to the assumption that the hot gas is located in the Galactic thick disk \citep[e.g.,][]{futamoto2004,yao2005,wang2005,yao2009,pinto2013}. This hot ISM derived from XRBs observations has an exponential scale height of several kpc \citep{yao2005,yao2009b}. The hot gas near the Galactic plane may has density around 1$\times$10$^{-3}$ cm$^{-3}$ \citep{yao2009x,hagihara2010,Sakai2014} and the temperature around 2 $\times$10$^{6}$ K \citep{yao2007}.  By comparing the amount of absorption detected along the sightline towards the XRBs vs. those detected toward the distant AGNs, it has been concluded that most of the hot gas is constrained within the disk \citep{yao2008}. 

However, the circumstellar medium local to XRB could also produced such absorption lines. \citet{miller2004} analyzed the absorption lines from GX~339--4 and XTE~J1650--500 and they find these highly ionized absorption lines could not be produced in the ISM, but in a photo-ionized warm absorber intrinsically. Later work on the time variability of the absorption lines show that in some targets the line strength varies between observations taken at different time \citep{cackett2008,luo2014,Petrucci2014}. These lines are likely photo-ionized by the XRB luminosity with an ionization parameter log($\xi$) in the range of 1 - 3 \citep{cackett2008,luo2014}. Some of the highly-ionized lines also are found to be blue-shifted, and could be modeled with outflowing photoionized gas near the accretion disk or its corona \citep[e.g, SAX~J1808.4--3658 observed by][]{pinto2014}. Study of the origin of these highly ionized absorption lines would benefit our understanding of both the circumstellar medium intrinsic to the XRB and the ISM, as well as the hot gas content in our Galaxy in general. 

In order to fully understand the origin of the absorption lines detected in the X-ray spectra of the XRBs and thereafter the properties of the hot ISM, a large sample of targets with detection of hot gas produced absorption lines is required. In this work, we perform a comprehensive analysis of the highly ionized absorption lines in the XRBs, with a focus on the highly ionized \ion{O}{7} absorption line at 21.6 \AA. We have performed a comprehensive analysis of all the available data in the {\sl XMM}-Newton archive with a concentration on \ion{O}{7} lines. 
Recent work by \citet{fang2015} provided an all-sky survey of \ion{O}{7} toward background active galactic nuclei (AGN) from archival XMM-Newton observations, and they find a 100\% sky covering fraction of \ion{O}{7} lines, which suggest a uniform distribution of the absorbers. Here we present the second paper in this series.

The paper is organized as follows. In section 2, we first describe the selection of our sample and the procedure of data reduction, and then we analyze the absorption lines and present our main results in this section. We discuss the distribution of the absorption line in section 3, as well as the origins of these lines. We also comment on several targets with newly detected, \ion{O}{7} lines in this section. The last section is the summary.

\section{Observation and Data Analysis}
\subsection{Target Selection and Data Reduction} 

\begin{figure*}
\center
\includegraphics[width=.75\textwidth,height=.25\textheight,angle=0]{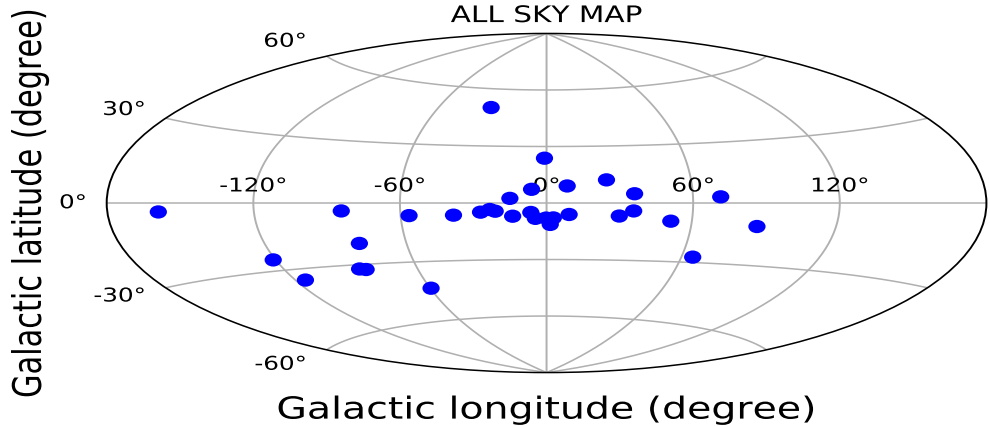}
\caption{All-sky Hammer-Aitoff projection of targets in our sample.}
\label{fig:aitoff}
\end{figure*}

Numerous X-ray absorption lines were detected in the X-ray spectra of the XRBs. These absorption lines sample the ISM in a variety of temperature ranges, providing rich information of the multi-phase distribution of the ISM \citep[e.g.,][]{ferriere2001,yao2005,costantini2012,pinto2013}. Here we focus on highly ionized, He-like \ion{O}{7} for the following reasons. First, as one of the most abundance elements in universe, oxygen is a very important tracer of the metals in the ISM. Secondly, under collisional ionization, the peak temperature of \ion{O}{7} ionization fraction is in the range of 10$^{5.5}$-10$^{6.5}$ K, providing an effective way to probe the hot, ionized ISM. Therefore, in our work, we focused on the absorption line from \ion{O}{7} K$_{\alpha}$ transition with a rest-frame wavelength of $\lambda_{rest}= 21.6019\rm\ \AA$ \citep{verner1996,yao2009,liao2013}.

Currently, there are three grating-based, high resolution X-ray spectrometers suitable to our study of highly ionized metal absorption lines, namely the Low and High Energy Transmission Grating Spectrometer\footnote{see {\sl Chandra} Proposers' Observatory Guide at: http://cxc.harvard.edu/proposer/POG/html/.} (LETG and HETG) onboard the {\sl Chandra} X-ray Observatory and the Reflection Grating Spectrometer\footnote{see XMM-Newton Users' Handbook at: https://xmm-tools.cosmos.esa.int/external/xmm\_user\_support/documentation/uhb/.} (RGS) onboard {\sl XMM}-Newton X-ray telescope. RGS has relatively high collecting area than both LETG and HETG. For RGS, there are two units, RGS1 and RGS2. However, one CCD assembly in each RGS unit has an operation failure due to electronics problems. These are RGS1 CCD7 and RGS2 CCD4, roughly covering the wavelength ranges 11-14 \AA\ and 20-24 \AA\, respectively, in first order. Therefore, in this work, we will focus on the RGS1 data only. 

Based on {\sl Chandra} observation, \citet{luo2014} has already studied a sample of \ion{Ne}{9} line toward XRBs. Due to the small effective area and strong galactic absorption, the detected \ion{O}{7} photon counts from {\sl Chandra} are small. In our selection of all XRBs observations, the number of XRBs observed by {\sl Chandra} is smaller than that of RGS, therefore we could not make a large well-defined \ion{O}{7} sample from {\sl Chandra}. At wavelength of 19.0\AA\ the RGS1 has one instrumental feature which make it unlikely for a good fit of \ion{O}{8} K$_\alpha$ line. At wavelength of 18.6\AA\, we found some weak detections of \ion{O}{7} K$_\beta$ line in part of our sample. Due to the small line oscillation strength, the line EWs and their significances could not be well constrained. We have tried to jointly fit the \ion{O}{7} K$_\alpha$ and K$_\beta$, and the joint-fit could not give a better improvement of the line EWs. Therefor in our work, we will concentrate on the \ion{O}{7} K$_{\alpha}$ line from RGS1 observation.

We have compiled a list of XRBs from the {\sl XMM}-Newton archive. We inspected the RGS data of these targets and selected the ones with relatively high photon statistics. We define our selection criteria as the counts per resolution element (CPRE) at wavelength near \ion{O}{7} must be at least 20 photos. For each source, we define the CPRE, which is the photon counts within a bin size of the spectral resolution (50 m\AA):
\begin{equation}
CPRE = F_\lambda A T \Delta\lambda,
\end{equation}
where $F_\lambda$ is the photon flux, $A$ is the effective area of the detector, $T$ is the observation time, and $\Delta\lambda$ is the width of one resolution element. We also ignore those targets with \ion{O}{7} K-shell emission lines (consisting of resonance, inter-combination and forbidden lines at 21.60, 21.80 and 22.10 \AA, respectively) to avoid contamination. This leads to a sample of 33 sources, with 29 low mass X-ray binaries (LMXBs) and 4 high mass X-ray binaries (HMXBs). In Figure \ref{fig:aitoff} we plot the all-sky Hammer-Aitoff projection of our targets in the Galactic coordinates. Most of our targets are located in or near the Galactic disk. 

We have reduced the spectra with the Science Analysis System (SAS) version 13.0 \footnote{see Cookbook at: http://heasarc.gsfc.nasa.gov/docs/xmm/abc/.}. RGS data are processed with the SAS task {\sl rgsproc}. Events with flags of $BAD\_SHAPE$, $ON\_BADPIX$, $ON\_WINDOW\_BORADER$, and $BELOW\_ACCEPTANCE$ are all rejected. We also did not keep the cool pixels (keepcool=no). We produced the light curves for the background in CCD9 following the XMM-SAS guide in order to remove time intervals that are contaminated with soft proton flares and spurious events. For each observation, only the first order spectrum was extracted.  

\subsection{Absorption Line Analysis}

Many of these sources were observed multiple times. To avoid unreal artifacts that may be introduced in the co-added exposures, for each target we simultaneously fitted all the exposures, each with its own response. Our analysis focuses on the segment spectrum around 20 -- 22 \AA\ and models the spectra continuum with the Galactic neutral absorption plus a power-law model, using XSPEC ver 12.7 \citep{Arnaud1996}. For each target, while we have adopted different continuum for each exposure, the parameters of the interstellar absorption were always tied together. We also excluded wavelength regions of known detector features. Absorption lines were fitted with a Voigt line profile model developed in \citet{buote2009}. This line model has three free parameters: Column density, Doppler-b parameter, and the velocity shift of the central wavelength. We refer the reader to \citet{buote2009} and \cite{fang2010} for details. Since we are interested in the absorption lines produced by the local hot gas, therefore we limited the line shift-velocity within 500 km s$^{-1}$. The Doppler-b parameter is also limited in the range of 20 to 300 km s$^{-1}$. The lowest temperature that can still produce a significant fraction of \ion{O}{7} is around 5 $\times$ 10$^5$ K under collisional ionization equilibrium, this corresponds to a thermal velocity of $\sim$ 20 km s$^{-1}$. The upper limit is adopted by assuming the absorbing gas shall not escape from our Galaxy. For a better constraint of the line EWs, we estimate the statistical significance of the lines using Monte Carlo simulations. Briefly, we make a simulated spectra based on our fitted absorption line models and run 1,000 simulations on each targets. We fit the simulated spectra, record the measured EWs, and finally obtain the line significance \citep[more details see][]{fang2010}. We performed the fit by minimizing the C-statistic, which yields less biased best-fitting parameters.

The basic information of our sample and the fitted line properties are listed in Table \ref{table:result}. We show the source name and type in column 1 and 2, respectively. Column 3 is the galactic neutral hydrogen column density adopted from \citet{kalberla2005}. Columns 4, 5, 6 are the galactic latitude, longitude and distance, respectively. All of the distance are adopted from literatures. We list the total exposure time and CPRE for each source in column 7 and 8, respectively. Column 9, 10, 11 are the \ion{O}{7} column density, velocity shift of the line center and the Doppler-b parameter, respectively. The line equivalent width (EW) and 1$\sigma$ statistical uncertainty are given in column 12. We show the line significance in column 13 and $C$-statistic and degree of freedom in column 14. The targets are listed in descending order based on their CPRE, which indicates of the the quality of the spectra. For targets we cannot constrain either the Doppler-b parameter or the shift-velocity of the line center, we set the velocity shift of the line center at 0 km s$^{-1}$ or the Doppler-b parameter at 300 km s$^{-1}$. In our sample, we find one target having detected absorption line with shift velocity larger than 500 km s$^{-1}$. Such a large shift suggests most likely the absorption line is produced by the circumstellar medium local to the source. We also list this high velocity line in the table, plus with the 3$\sigma$ upper limits of EW for their lines at 0 km s$^{-1}$.  We adopt 1$\sigma$ errors throughout the paper unless otherwise mentioned.

\begin{deluxetable}{lllll}
\tablewidth{0pt}
\tabletypesize{\scriptsize}
\tiny
\tablecaption{Target Detection}
\tablehead{  & $> 3 \sigma$ & $> 2 \sigma$ & $> 1 \sigma$ & Total}            
\startdata                    
All       & 16 & 25 & 32 & 33\\
LMXB      & 14 & 22 & 28 & 29\\
HMXB      & 2  & 3  & 4  & 4 
\enddata
\label{table:detection}
\tablecomments{Target detection in our line measurements. Columns 2, 3, 4 list the targets detected with a significance larger than 3, 2, 1$\sigma$, respectively. The last column is the total numbers of targets in each group.}
\end{deluxetable}

Of the total of 33 targets, 16 show an \ion{O}{7} detection significance larger than 3$\sigma$ threshold, 9 are at the 2-3$\sigma$ level. Details about the detections at different significance level are listed in Table \ref{table:detection}. Only one target has a weak absorption line presented as significance less than 1$\sigma$. The difference between HMXB and LMXB mostly comes from the mass of the companion star and the way accretion of matter occurs. The spectra of HMXB is sensitive to the properties of the companion wind. In our sample, we have four HMXBs, two (LMC~X--3 and Cygnus~X--1) of which show \ion{O}{7} detection significance larger than 3$\sigma$. 4U~0513--40 has shift-velocity larger than 500 km s$^{-1}$, and its detection significance is 2.9$\sigma$. In the spectrum of EXO~0748--676, \ion{O}{7} K-shell emission is clearly shown near 21.8 \AA\ \citep{cottam2002,van-peet2009}, whereas one absorption line at 21.6\AA\ is also detected here with a significance of 4.1$\sigma$. Of the targets with significance larger than 3$\sigma$, 4 targets are newly newly detections: Sco~X--1, Swift~J1753.5--0127, MAXI~J0556--332 and MAXI~J1910--057. 

For a more careful examination of the line significances, we have tested the detections by another method of Monte Carlo simulations. We first fit and simulate the continuum spectrum, and then we fit the simulated data with a continuum model and record the C-statistic. We add the absorption model, and record the change of the C-statistic. We have performed 10,000 simulations for three test targets to check how many times the changed C-statistic given by adding a \ion{O}{7} line provides a result better than the measurement. These three test targets, Sco~X--1, EXO~0748--676 and GS~1826--238 are selected due to their different statistics and we expect this small sample will provide a very interesting view on the overall significance of the total sample. We get the line significances for these three targets are $>$6$\sigma$, 2.8$\sigma$, and 3.6$\sigma$, respectively. These results are in general consistent with the significances we have obtained in the first Monte Carlo simulations.

\begin{figure}
\center
\includegraphics[width=.45\textwidth,height=.3\textheight,angle=0]{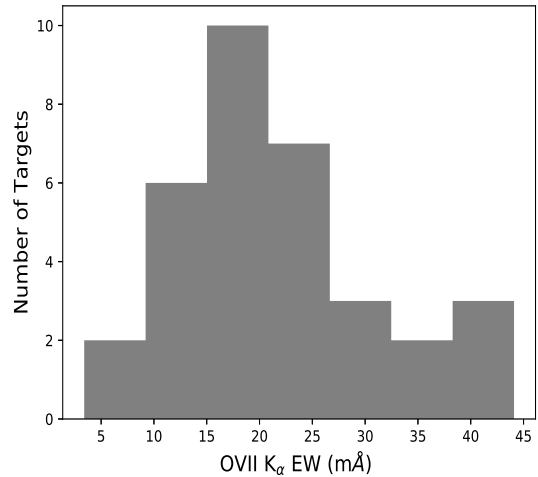}
\caption{Histogram distribution of the \ion{O}{7} K$_{\alpha}$ line EW at detected significance larger than 1$\sigma$. The targets have a EW centered around 20 m\AA\ and distribute in the range of 10 - 40 m\AA.}
\label{fig:hist}
\end{figure}

\begin{figure}
\center
\includegraphics[width=.45\textwidth,height=.3\textheight,angle=0]{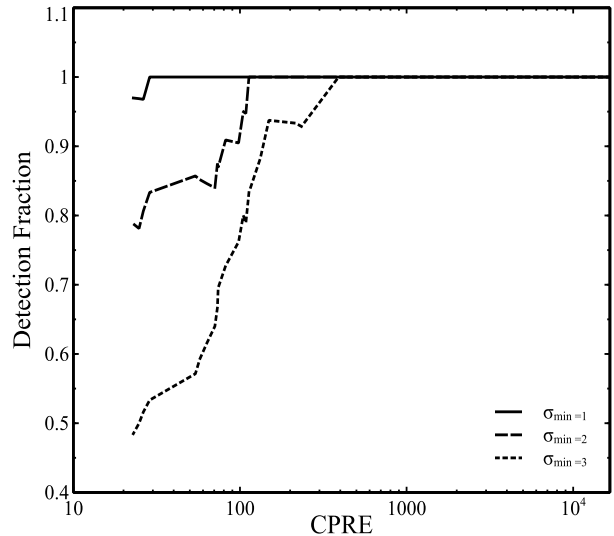}
\caption{Detection fraction as a function of source CPRE. The fractions for absorption lines detected with at least 1, 2, and 3$\sigma$ significance are presented as the solid, dashed, and dotted lines, respectively.}
\label{fig:cpre}
\end{figure}

In Figure \ref{fig:hist}, we show the histogram distribution of the \ion{O}{7} K$_{\alpha}$ line EW for targets detected at $>1 \sigma$ significance. The measured line EWs are centered around $\sim$ 20 m\AA, with a range of 10 - 40 m\AA. In Figure \ref{fig:cpre}, we show the line detection rate, or the sky covering fraction, as a function of the photon counts CPRE. The solid, dashed, and dotted lines are for the detections with at least 1, 2, and 3$\sigma$ significance, respectively. For all cases, the detection fractions increase steadily as the increasing CPRE. For targets with CPRE more than about 400 counts the 3$\sigma$ detection fraction reaches 100\%. Such detection rate implies a uniform distribution of the hot gas in all directions.

\begin{deluxetable}{llll}
\tablewidth{0pt}
\tabletypesize{\scriptsize}
\tiny
\tablecaption{Comparison with Previous Measurements}
\tablehead{ Name & This Work& {\sl Chandra} & {\sl XMM}-Newton \\
 & (m\AA) & (m\AA) & (m\AA) }            
\startdata                    
Cygnus~X--2	&	18.3 	$\pm$	1.5 &	$<$6.3			[1]		&	19.6	$\pm$	1.3	[2]		\\
XTE~J1817--330	&	34.2$\pm$	5.4 &	54.0	$\pm$	4.0	[4]	&	 26.4     $\pm$   7.6 [5]  		\\
4U~1820--30	&	20.5	$\pm$	2.1 &	40	$\pm$	7.9	[6]		&	23.3	$\pm$	2.5	[7]		\\
	         &					&		44.8 $\pm$ 10.8 [3]				&	23.9	$\pm$	3.6	[8]		\\
GX~339--4	&	42.5	$\pm$	3.5	&						&	23.7 $\pm$ 5.9 [9]\\ 
SAX~J1808.4--3658	&	18.5	$\pm$	2.5	&						&	31.5 $\pm$ 7.2 [9] \\
LMC~X--3	&	20.5	$\pm$	3.1	&	20	$\pm$	6	[10]		&	21.0	$\pm$	5.0	[8]		\\
4U~1636--54	&	27.7	$\pm$	4.5&						&	19.9 $\pm$ 7.0 [9] \\
4U~1728--16	&	10.1	$\pm$	4.7	&						&	8.6 $\pm$ 4.3 [9] \\ 
4U~1254--690	&	17.6	$\pm$	5.3	&						&	15.2 $\pm$ 7.6 [9]\\ 
Aql~X--1	&	25.0	$\pm$	8.6	&						&	18.1 $\pm$ 7.2 [9]\\ 
4U~1957+11	&	16.1	$\pm$	6.3	&	18	$\pm$	18	[11]		&	19.0	$\pm$	19.0	[11]		\\
	&		&	18.7	$\pm$	10.4	[12]		&						\\
GS~1826--238	&	31.6	$\pm$	10.0	&						&	18.1 $\pm$ 7.2 [9]\\
4U~1735--44	&	26.4 	$\pm$	11.3	&						&	19.9 $\pm$ 10.0 [9]\\ 
	&			&						&	24.7	$\pm$	9.7	[8]		\\
Ser~X--1	&	27.5	$\pm$	11.9&						&	21.0 $\pm$ 4.4 [9]
\enddata
\label{table:consist}
\tablecomments{
[1]\citet{yao2009}
[2]\citet{cabot2013}
[3]\citet{futamoto2004}
[4]\citet{gatuzz2013}
[5]\citet{sala2007}
[6]\citet{yao2006}
[7]\citet{costantini2012}
[8]\citet{miller2013}
[9]\citet{pinto2013}
[10]\citet{wang2005}
[11]\citet{nowak2008}
[12]\citet{yao2008}    
}
\end{deluxetable}

Some of the \ion{O}{7} detections were reported previously. We have compared our results with previous measurements. We list our work and previous studies in Table \ref{table:consist}. Column 2 shows the line EWs in our work, and columns 3 and 4 are previous results obtained with {\sl Chandra} and {\sl XMM}-Newton, respectively. Some of the previous results were initially listed in units of column density. For consistency, we have converted those results to units of m\AA\ with 1$\sigma$ error bar, assuming a Doppler-b parameter of 100 km s$^{-1}$. Our measurements are largely consistent with previous ones within the error range. Cygnus~X--2 is a unique one because the \ion{O}{7} K$\alpha$ line is not detected in {\sl Chandra}. We refer the reader to \citet{cabot2013} for a detailed analysis of the \ion{O}{7} line in Cygnus~X--2. For XTE~J1817--330, \citet{sala2007} obtained an \ion{O}{7} EW of 0.7 eV (26.4 m\AA) and is consistent with our measurements, whereas the {\sl Chandra} observation by \citet{gatuzz2013} gives a higher value. Similar discrepancy was also presented in 4U~1820--30. One explanation for the discrepancy may be the time variability of \ion{O}{7} line since {\sl Chandra} observations were performed in a different time. Part of our sample have been analyzed by \citet{pinto2013}. Our measurements are in general consistent with theirs. One exception is GX~339--4, for which we found a much higher EW. The difference may be caused the different data reduction procedure. For this target, its all observations were not co-added together as previous did but were simultaneously analyzed. Highly ionized absorption line in LMC~X--3 has been systematically studied by \citet{wang2005} using high resolution spectrometers on board {\sl Chandra} X-ray telescope, and our results are consistent with their measurements. In the {\sl Chandra} spectra of Cygnus~X--1, \citet{schulz2002} find an a weak detection at 21.43 \AA\ and they identify this line as an blue-shift \ion{O}{7} K$\alpha$ line, while the blueshifted velocity is detected differently as 2350 km s$^{-1}$ and we get the velocity as 494 km s$^{-1}$. This could be a signature of line variation. 

\section{Discussion}

\subsection{Study of ISM Distribution}

\begin{figure*}
\center
\includegraphics[width=\textwidth,height=.35\textheight,angle=0]{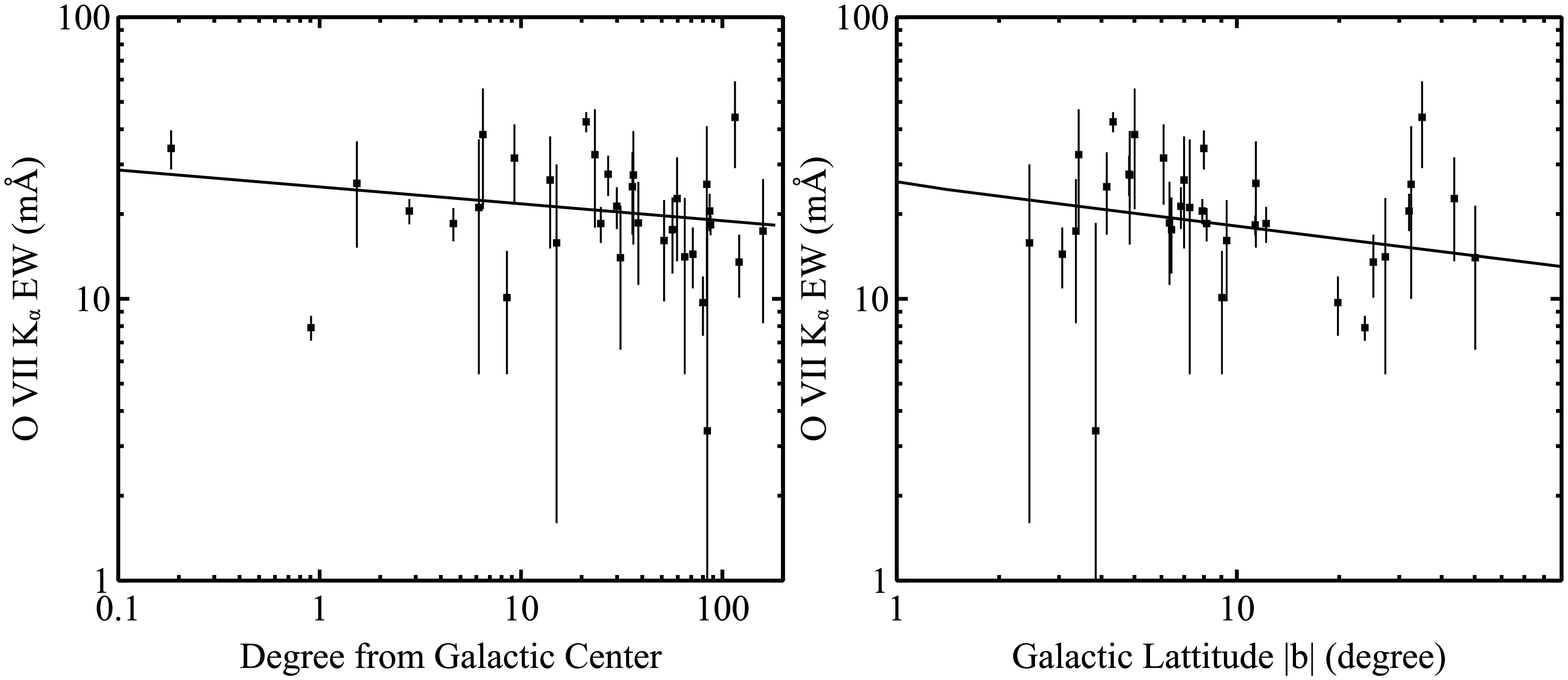}
\caption{\ion{O}{7} K$_{\alpha}$ EW against the galactic longitude (left panel) and latitude (right panel).}
\label{fig:ew2lb}
\end{figure*}

\begin{figure}
\center
\includegraphics[width=.45\textwidth,height=.3\textheight,angle=0]{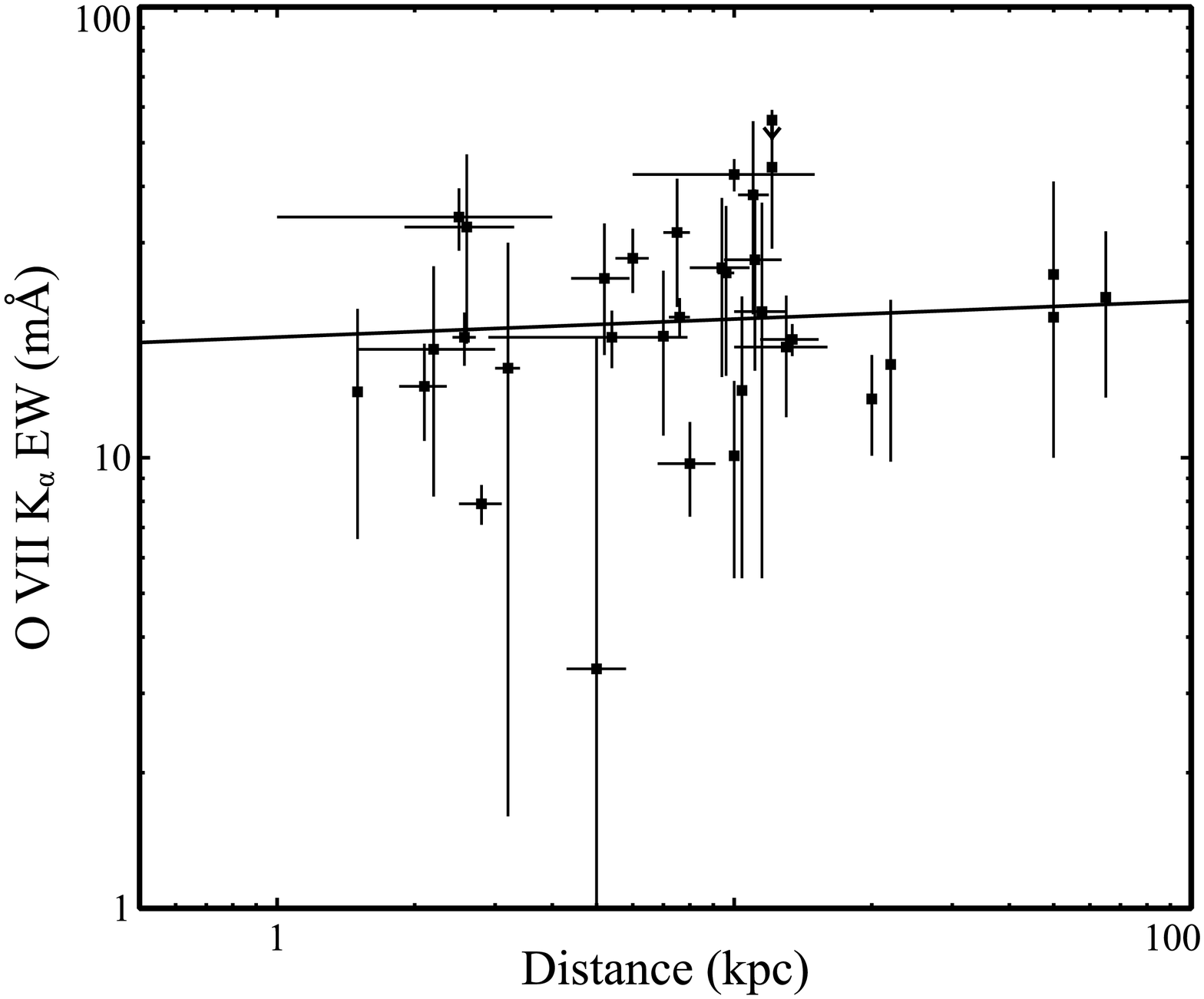}
\caption{\ion{O}{7} K$_{\alpha}$ EW as function of distance.}
\label{fig:dis2ew}
\end{figure}

\begin{figure}
\center
\includegraphics[width=.45\textwidth,height=.3\textheight,angle=0]{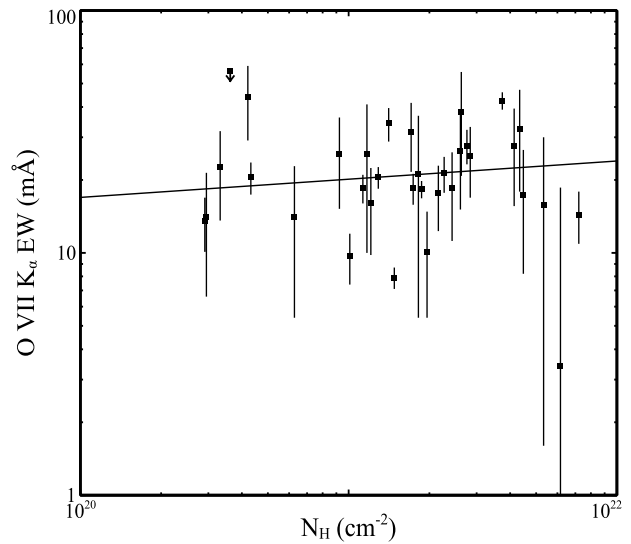}
\caption{EWs of \ion{O}{7} K$_{\alpha}$ line compared to galactic hydrogen absorption N$_{\rm H}$. The N$_{\rm H}$ is adopted from \citet{kalberla2005}}.
\label{fig:ew2nh}
\end{figure}

The distribution of the metal absorption lines in the spectra of different XRBs allows us to quantify how the intervening hot gas distributed in the circumstellar medium and the ISM. Here we show the dependance of the line EW on the target direction, neutral gas column density and distance. In Figure \ref{fig:ew2lb}, we show the relations between \ion{O}{7} K$_{\alpha}$ EW and galactic latitude (left panel) and longitude (right panel). In the left panel, for $l$ in the range of 50  to 180 degree, the mean EW is less than  20 m\AA, while toward the Galactic center for $l$ in the range of 0 to 30 degree, the mean EW is larger than 20 m\AA. For the right panel, a large part of the sample has coordinates toward the galactic center with coordinates $|b|$ in the range of 0 to 10 degree, however, the tendency for the EWs along with $b$ is rather weak. At higer latitude $|b| > 20$, the EWs slightly increase with $|b|$, while at lower latitude, there is a large scatter of EWs.

In Figure \ref{fig:ew2nh} we show the measured \ion{O}{7} K$_{\alpha}$ EW as a function of the Galactic hydrogen absorption N$_{\rm H}$. It could be found the EWs likely show little variation with N$_{\rm H}$ and there is no clear correlation between the EWs and N$_{\rm H}$. Assuming a mean \ion{O}{7} EW of 20 m\AA\ and a Doppler-b parameter of 100 km s$^{-1}$, we could obtain an \ion{O}{7} column density of about 2$\times$10$^{16}$ cm$^{-2}$. Assuming solar abundance and all oxygen of the hot gas would be in the ionization stage of \ion{O}{7}, the total hydrogen column density of the hot gas would be 4.1$\times$10$^{19}$ cm$^{-2}$. By taking the ratio of the hot gas and neutral gas column density, we could find the relative fraction with neutral hydrogen would be in the range of 0.4 to 0.004. 

To investigate how the hot ISM distributes around the disk, we show the \ion{O}{7} K$_{\alpha}$ EW vs. the target distance in Figure \ref{fig:dis2ew}. In general the EWs of the targets show a very weak and positive correlation, and also show scatters by a factor of a few or more among targets at similar distances. For targets with distance larger than 10 kpc, their EWs are mostly greater than 10 m\AA, and with distance larger than 50 kpc, their EWs are greater than 20 m\AA. While for targets with distance less than 5 kpc, there are at least 2 objects with EW $>$ 30 m\AA. However, due to large scattering, the trend between EW and distance is not obvious, and at small distance this large scatters coule be a result of the intrinsic absorption contamination.

We have calculated the dependance of the line EW on the galactic neutral absorption, target distance and the galactic latitude and longitude. Briefly, we first use software package ASURV Rev 1.2 \citep{Isobe1990,Lavalley1992}, which implements the methods presented in \citet{Isobe1986} and is a particular useful tool for correlation analysis of censored astronomical data, i.e., non-detections or detection limits. Then we characterize a correlation with a simple log-log linear relation, together with the dispersion around this relation from the root mean square (RMS) of the data. The corresponding RMS could give the probabilistic nature of the correlation and the measurement uncertainties. We listed the probability of correlation by chance for three correlation tests: Cox hazard model, generalized Kendall's tau, and Spearman's rho in Table \ref{table:correlation} column 1, 2, and 3, respectively. A correlation exists if the probability is less than 5\%. In general, the probability for all correlations are much larger than 5\%. In column 4 and 5, we list the index of the fitted log-log linear relation and the corresponding RMS, respectively. The index is quite close to zero, and the correlations are significantly sub-linear. With all these in consideration, we may then conclude that EW has no correlation with the galactic neutral absorption, the target distance or the galactic coordinates.

If the hot gas probed by the \ion{O}{7} absorption line is uniformly distributed in the ISM, one would expect the detected \ion{O}{7} column density may show some correlations with physical properties such as the the distance and/or coordinates of the background targets. Assuming the absorption line is in the linear part of the curve-of-growth, such correlations can be presented as the correlations between the line EW and these physical quantities. The non-detection of the correlations may suggest the line is in the non-linear part of the curve-of-growth and we do not have strong constraint on the line Doppler-b parameter. However, it is unlikely caused by the a non-uniform distribution of the hot gas properties, such as temperature, metallicity, or density, because of uniform distribution of the \ion{O}{7} absorption lines. Besides of these, the contamination of the circumstellar material, and the historical supernova activity \citep{ferriere2001} could also introduce absorption to the lines.

One example is studying spatially closed targets for which the ISM distribution along the sightline towards these targets should be similar. We searched our sample for targets close to each other by their relative $l$ and $b$ within 5 degree. First we find that two targets SAX~J1808.4--3658 and 4U~1820--30 show comparable \ion{O}{7} EW, whereas their distance differ by three times. The less distant target XTE~J1817--330 has EW much larger than these two targets even though all three targets are spatial close to each other. Similarly in another group the EW for 4U~1636--54, with error bar taking into consideration, is comparable with that of XTE~J1650--500, while 4U~1636--54 has a distance twice far away. Even though there are still too few pairs to draw definitive conclusions, this result may suggest that in some cases the intrinsic absorption dominates the ISM contribution.

\begin{deluxetable*}{llccccc}
\tablewidth{0pt}
\tabletypesize{\scriptsize}
\tiny
\tablecaption{Correlation Test}
\tablehead{ Correlation & Cox Hazard & Kendall Tau & Spearman Rho & $\Gamma$ & RMS (dex)}           
\startdata      
EW vs. N$_{\rm H}$  & 40\% & 87\% & 87\% & 0.075$\pm$0.001 & 0.23$\pm$0.06\\
\hline
EW vs. Distance  & 20\% & 21\% & 19\% & 0.04$\pm$0.12 & 0.26$\pm$0.07\\
\hline
EW vs. $l$  & 36\% & 32\% & 34\% & -0.06$\pm$0.08 & 0.26$\pm$0.07 \\
\hline
EW vs. $|b|$  & 53\% & 40\% & 37\% & -0.15$\pm$0.18 &0.26$\pm$0.08
\enddata
\label{table:correlation}
\end{deluxetable*}

\subsection{Comments on Some Individual Targets}

We briefly discuss several individual targets with newly detected \ion{O}{7} absorption lines at more than 3$\sigma$ significance level. In our sample, we find 16 targets showing line significance at more than 3$\sigma$ level, of which 4 targets are newly detected, Sco~X--1, Swift~J1753.5--0127, MAXI~J0556--332 and MAXI~J1910--057.

Sco~X--1 -- This target is one of the brightest X-ray sources in the sky. With a high signal-to-noise ratio spectrum we can achieve much better constraints of the absorption line properties. Previous studies based on absorption lines mostly focus on the low ionization absorption lines or absorption edges, and our measurement is the first report on the \ion{O}{7} detection \citep{Garcia2011,de-Vries2009}. The \ion{O}{7} line in Sco~X-1 is detected at 9.9$\sigma$ significance level and the line shift velocity is very low (less than 100 km s$^{-1}$).

Swift~J1753.5--0127 -- This is also a LMXB discovered with the {\sl Swift}/BAT on May,2005 \citep{palmer2005}. \citet{mostafa2013} report the discovery of broad emission lines of \ion{N}{7} and \ion{O}{8} in the RGS spectrum. They attributed this feature as a result of reflection of X-ray photons off accretion disk. 

MAXI~J0556--332 --  This is a X-ray transient. \citet{maitra2011} reported the detection of a strong emission line near 24.8\AA, identified as \ion{N}{7}. So far there is no report of absorption lines produced by an outflow. We detect the \ion{O}{7} absorption with 4$\sigma$ significance level.

MAXI~J1910--057 -- The source was first detected by the {\sl Swift}/BAT \citep{krimm2012} on May, 2012 and simultaneously by the {\sl MAXI} telescope \citep{usui2012}. The observed spectra are typical of an LMXB \cite{charles2012} indicating that it is comparable with other neutron star/black hole soft X-ray transients, but the distance is still unclear yet. The \ion{O}{7} line in this target is first reported here with a blueshift velocity of 205 km s$^{-1}$, an EW of 21.3 m\AA\ and a significance of 6.0. 

4U~0513--40 -- This is an ultracompact X-ray binary in the globular cluster NGC 1851. This target was only observed by {\sl XMM}-Newton once in April, 2003. Previous X-ray observations did not report any detection of \ion{O}{7} absorption line and in our work we found this line having shift-velocity of -595 km s$^{-1}$ and detection significance at 2.9$\sigma$ level. Bt its high shift velocity, this absorption line could be a good indicator of an circumstellar medium origin.

\section{Summary}

Understanding the properties of the hot ionized medium has profound impact on the study stellar feedback and the multiphase structure of the ISM. In this work, we perform a survey in the {\sl XMM}-Newton archive for searching \ion{O}{7} absorption lines in the spectra of Galactic X-ray sources. We summarize our findings here.

\begin{itemize}

\item We analyzed 33 Galactic XRBS with 29 targets are LMXBs and 4 targets are HMXBs. Most targets are located at a distance if within 20 kpc, in the Galactic disk. 16 targets have \ion{O}{7} detection at more than 3$\sigma$, among which 4 are newly discovered in this work. We find one targets showing shift-velocity larger than 500 km s$^{-1}$ indicating of an circumsteller medium origin.

\item We fitted the \ion{O}{7} K$_\alpha$ transitions with a Voigt-profile based line model. We find that most K$_\alpha$ lines have an EW of 20 m\AA, with a range of 10-40 m\AA.

\item We find that the detection fractions, or the sky covering fractions, increase steadily as the increasing CPRE. For targets with CPRE more than about 400 counts the 3$\sigma$ detection fraction reaches 100\% regardless of the direction of the XRBs. Such detection rate implies a uniform distribution of the hot gas in all directions.

\item The EWs of the \ion{O}{7} line do not correlate with the galactic neutral absorption N$_{\rm H}$. The fraction of the content of hot medium to that of neutral phase is estimated in the range of 0.004 to 0.4. We do not find any strong correlations between \ion{O}{7} EW and target distance, so as to the targets coordinates. The EW for targets at similar distance show large scatters and the difference could be as large as one order of magnitude. The reasons for the non-correlation could be the contribution of circumstellar material or the uncertainties on the column density measurements, as we do not have constraints on the Doppler-b parameter, so EWs may not exactly reflect the column density distribution.

\end{itemize}

The highly ionized, z = 0 \ion{O}{7} absorption lines detected in the X-ray spectra of background AGNs, along with the detections in Galactic XRBs could help to constrain the distribution of hot ISM with a broad range of temperature around 10$^6$ K. We briefly compare our results with the AGN \ion{O}{7} line survey by \citet{fang2015}, and we find there are some similarities of the line properties. The detection fractions, in both surveys, increase steadily as the increasing CPRE, which suggesting a uniform distribution of the \ion{O}{7} absorbers. The histogram of K$_\alpha$ lines in both cases have an median EW of 20 m\AA, and there are also no correlations between the EWs and the targets coordinates. 

By considering the different coordinates and distances or covering area and absorbing depth of the AGNs and XRBs, we could find that the XRBs have a large fraction of contaminations intrinsically contributing to the \ion{O}{7} absorption. But how much the XRBs locally contribute to the absorption still needs more investigations. Further theoretical modeling in combine with X-ray emission study, and X-ray absorption study will be presented in a later paper.

\acknowledgments
We thank Drs. David Buote and Daniel Wang for helpful discussion, This work is supported by National Key R$\&$D Program of China No. 2017YFA0402600, NSFC grants No. 11525312, 11333004, 11443009, and U1531130. TF is also supported by the Specialized Research Fund for the Doctoral Program of Higher Education (SRFDP; No. 20130121110009).

\bibliography{main}

\clearpage
\clearpage
\begin{landscape}
\begin{deluxetable}{llllllllllllll}
\tablewidth{0pt}
\small
\tablecaption{The X-ray Targets Sample}
\tablehead{ Source & Type& N$_{\rm H}$  &l & b & Distance & Exp. & Counts & Log(N$_{\rm OVII}$) & Velocity & Doppler-b & EW(K$_\alpha$)  &  S/N & $C$/dof \\
 & & (cm$^{-2}$) & & & (kpc) & (ksec) & & (cm$^{-2})$ & (km s$^{-1}$) & (km s$^{-1}$) & (m\AA) & ($\sigma$) & }            
\startdata                    
Sco~X--1	&	LMXB        	&	1.48E+21	&	359.094	&	23.784	&	2.8	$_{	-0.3	}^{+	0.3	}$	[6]	&	36.8	&	16612.7	&	15.8	$^{	+1.2	}_{	-0.4	}$&	15.4	$^{	+98.9	}_{	-93.4	}$&	42.8	$^{	+274.2	}_{	-22.8	}$&	7.9	$\pm$	0.8	&	9.9	&	972/662	\\
Cygnus~X--2	&	LMXB        	&	1.87E+21	&	87.328	&	-11.316	&	13.4	$_{	-2	}^{+	1.9	}$	[6]	&	199.9	&	3089.7	&	16.3	$^{	+1.2	}_{	-0.5	}$&	145.5	$^{	+108.4	}_{	-108.3	}$&	78.9	$^{	+217.4	}_{	-50.9	}$&	18.3	$\pm$	1.5	&	12.2	&	1190/971	\\
XTE~J1817--330	&	LMXB        	&	1.41E+21	&	359.817	&	-7.996	&	2.5	$_{	-1.5	}^{+	1.5	}$	[6]	&	18.0	&	1379.4	&	16.2	$^{	+0.2	}_{	-0.2	}$&	2.3	$^{	+271.3	}_{	-304.2	}$&	300.0	$^{	+0.0	}_{	-300.0	}$&	34.2	$\pm$	5.4	&	6.3	&	219/206	\\
4U~1820--30	&	LMXB        	&	1.29E+21	&	2.788	&	-7.914	&	7.6	$_{	-0.4	}^{+	0.4	}$	[24]	&	76.4	&	1185.2	&	16.8	$^{	+0.9	}_{	-0.9	}$&	56.4	$^{	+108.0	}_{	-110.3	}$&	75.4	$^{	+224.6	}_{	-35.2	}$&	20.5	$\pm$	2.1	&	9.8	&	347/327	\\
GX~339--4	&	LMXB        	&	3.74E+21	&	338.939	&	-4.326	&	10	$_{	-4	}^{+	5	}$	[1]	&	443.1	&	1017.9	&	16.7	$^{	+0.1	}_{	-0.1	}$&	-71.3	$^{	+81.5	}_{	-80.3	}$&	202.2	$^{	+60.7	}_{	-43.3	}$&	42.5	$\pm$	3.5	&	12.1	&	1384/1354	\\
SAX~J1808.4--3658	&	LMXB        	&	1.13E+21	&	355.385	&	-8.148	&	2.57	$_{	-0.15	}^{+	0.15	}$	[2]	&	58.8	&	910.6	&	15.9	$^{	+1.6	}_{	-0.1	}$&	-118.1	$^{	+186.5	}_{	-216.3	}$&	300.0	$^{	+0.0	}_{	-280.0	}$&	18.5	$\pm$	2.5	&	7.4	&	289/206	\\
Swift~J1753.5--0127	&	LMXB        	&	1.74E+21	&	24.898	&	12.186	&	5.4	$_{	-2.5	}^{+	2.5	}$	[11]	&	118.1	&	798.8	&	15.9	$^{	+0.1	}_{	-0.2	}$&	-185.4	$^{	+220.2	}_{	-259.9	}$&	300.0	$^{	+0.0	}_{	-280.0	}$&	18.5	$\pm$	2.7	&	6.9	&	572/520	\\
EXO~0748--676	&	LMXB        	&	1.01E+21	&	279.975	&	-19.811	&	8	$_{	-1.2	}^{+	1.1	}$	[15]	&	382.9	&	691.8	&	16.0	$^{	+1.7	}_{	-0.7	}$&	24.5	$^{	+204.2	}_{	-170.5	}$&	46.2	$^{	+253.8	}_{	-26.2	}$&	9.7	$\pm$	2.3	&	4.2	&	1877/1433	\\
Cygnus~X--1	&	HMXB        	&	7.21E+21	&	71.335	&	3.067	&	2.1	$_{	-0.25	}^{+	0.25	}$	[16]	&	194.4	&	595.0	&	15.8	$^{	+1.7	}_{	-0.3	}$&	-494.5	$^{	+379.6	}_{	-5.5	}$&	251.4	$^{	+48.6	}_{	-231.4	}$&	14.4	$\pm$	3.5	&	4.1	&	1629/1472	\\
MAXI~J0556--332	&	LMXB        	&	2.91E+20	&	238.940	&	-25.183	&	20	$_{		}^{		}$	[5]	&	71.7	&	576.4	&	16.1	$^{	+1.6	}_{	-0.7	}$&	84.6	$^{	+259.5	}_{	-274.6	}$&	68.2	$^{	+231.8	}_{	-48.2	}$&	13.5	$\pm$	3.4	&	4.0	&	322/369	\\
LMC~X--3	&	HMXB        	&	4.32E+20	&	273.576	&	-32.082	&	49.97	$_{		}^{		}$	[17]	&	146.1	&	564.0	&	16.5	$^{	+1.2	}_{	-0.7	}$&	92.9	$^{	+139.5	}_{	-137.2	}$&	85.6	$^{	+214.4	}_{	-46.5	}$&	20.5	$\pm$	3.1	&	6.6	&	960/955	\\
MAXI~J1910--057	&	LMXB        	&	2.27E+21	&	29.903	&	-6.844	&	-	$_{		}^{		}$		&	51.2	&	529.1	&	15.9	$^{	+1.5	}_{	-0.2	}$&	-205.4	$^{	+254.5	}_{	-250.2	}$&	300.0	$^{	+0.0	}_{	-271.4	}$&	21.3	$\pm$	3.6	&	5.9	& 	225/120	\\
4U~1636--54	&	LMXB        	&	2.76E+21	&	332.915	&	-4.818	&	6	$_{	-0.5	}^{+	0.5	}$	[9]	&	245.1	&	390.9	&	16.7	$^{	+1.0	}_{	-0.7	}$&	-170.8	$^{	+171.0	}_{	-164.4	}$&	113.0	$^{	+187.0	}_{	-59.2	}$&	27.7	$\pm$	4.5	&	6.2	&	1064/1067	\\
4U~1728--16	&	LMXB        	&	1.96E+21	&	8.513	&	9.038	&	10	$_{		}^{		}$	[10]	&	32.1	&	234.0	&	15.9	$^{	+1.6	}_{	-0.9	}$&	-364.7	$^{	+417.8	}_{	-135.3	}$&	55.1	$^{	+244.9	}_{	-35.1	}$&	10.1	$\pm$	4.7	&	2.1	&	452/401	\\
4U~1254--690	&	LMXB        	&	2.16E+21	&	303.482	&	-6.424	&	13	$_{	-3	}^{+	3	}$	[13]	&	214.0	&	215.9	&	15.9	$^{	+1.8	}_{	-0.4	}$&	-35.3	$^{	+416.9	}_{	-340.4	}$&	203.5	$^{	+96.5	}_{	-183.5	}$&	17.6	$\pm$	5.3	&	3.3	&	685/664	\\
Aql~X--1	&	LMXB        	&	2.84E+21	&	35.718	&	-4.143	&	5.2	$_{	-0.8	}^{+	0.7	}$	[6]	&	95.1	&	149.1	&	16.0	$^{	+1.4	}_{	-0.3	}$&	248.4	$^{	+251.6	}_{	-520.0	}$&	299.4	$^{	+0.6	}_{	-279.4	}$&	25.0	$\pm$	8.1	&	3.1	&	405/386	\\
4U~1957+11	&	LMXB        	&	1.21E+21	&	51.309	&	-9.331	&	22	$_{		}^{		}$	[21]	&	44.8	&	132.4	&	16.4	$^{	+1.3	}_{	-1.0	}$&	-281.5	$^{	+584.1	}_{	-218.5	}$&	69.5	$^{	+230.5	}_{	-49.5	}$&	16.1	$\pm$	6.3	&	2.6	&	164/162	\\
4U~1543--62	&	LMXB        	&	2.43E+21	&	321.755	&	-6.337	&	7	$_{		}^{		}$	[3]	&	49.1	&	113.2	&	16.5	$^{	+1.0	}_{	-1.1	}$&	-189.3	$^{	+307.9	}_{	-310.7	}$&	71.3	$^{	+228.7	}_{	-51.3	}$&	18.6	$\pm$	7.4	&	2.5	&	190/207	\\
Swift~J1357.2--0933	&	LMXB           	&	2.95E+20	&	328.839	&	50.210	&	1.5	$_{		}^{		}$	[19]	&	27.8	&	108.5	&	16.0	$^{	+1.7	}_{	-1.0	}$&	48.5	$^{	+451.5	}_{	-548.5	}$&	90.8	$^{	+209.2	}_{	-70.8	}$&	14.0	$\pm$	7.4	&	1.9	&	153/163	\\
GS~1826--238	&	LMXB        	&	1.71E+21	&	9.272	&	-6.088	&	7.5	$_{	-0.5	}^{+	0.5	}$	[6]	&	228.2	&	104.7	&	16.2	$^{	+0.3	}_{	-0.4	}$&	-437.4	$^{	+535.8	}_{	-62.6	}$&	300.0	$^{	+0.0	}_{	-280.0	}$&	31.6	$\pm$	10.0	&	3.2	&	290/309	\\
4U~2129+12	&	LMXB        	&	6.28E+20	&	65.013	&	-27.312	&	10.4	$_{		}^{		}$	[7]	&	86.4	&	98.1	&	16.4	$^{	+1.0	}_{	-1.4	}$&	-133.4	$^{	+433.4	}_{	-366.6	}$&	42.4	$^{	+257.6	}_{	-22.4	}$&	14.1	$\pm$	8.7	&	1.6	&	297/242	\\
4U~1735--44	&	LMXB        	&	2.61E+21	&	346.054	&	-6.994	&	9.4	$_{	-1.4	}^{+	1.4	}$	[14]	&	20.1	&	82.0	&	16.1	$^{	+0.3	}_{	-0.4	}$&	-75.0	$^{	+475.0	}_{	-345.8	}$&	300.0	$^{	+0.0	}_{	-193.7	}$&	26.4	$\pm$	11.3	&	2.3	&	170/163	\\
4U~0614+09	&	LMXB        	&	4.48E+21	&	200.877	&	-3.364	&	2.2	$_{	-0.7	}^{+	0.8	}$	[20]	&	23.4	&	74.0	&	16.5	$^{	+1.2	}_{	-1.5	}$&	206.8	$^{	+293.3	}_{	-572.2	}$&	72.3	$^{	+227.7	}_{	-52.3	}$&	17.4	$\pm$	9.2	&	1.9	&	232/229	\\
SMC~X--1	&	HMXB        	&	3.32E+20	&	300.415	&	-43.559	&	65	$_{		}^{		}$	[23]	&	36.5	&	73.2	&	16.5	$^{	+1.2	}_{	-0.8	}$&	318.8	$^{	+181.2	}_{	-571.4	}$&	101.6	$^{	+198.4	}_{	-81.6	}$&	22.7	$\pm$	9.1	&	2.5	&	209/199	\\
GRO~J1655--40	&	LMXB        	&	5.33E+21	&	344.982	&	2.456	&	3.2	$_{	-0.2	}^{+	0.2	}$	[6]	&	113.8	&	70.7	&	15.8	$^{	+1.7	}_{	-0.8	}$ &	-230.5	$^{	+297.2	}_{	-269.5	}$&	61.1	$^{	+238.9	}_{	-41.1	}$&	15.8	$\pm$	14.2	&	1.1	&	351/331	\\
XTE~J1650--500	&	LMXB        	&	4.34E+21	&	336.718	&	-3.427	&	2.6	$_{	-0.7	}^{+	0.7	}$	[17]	&	23.9	&	63.2	&	16.1	$^{	+1.6	}_{	-1.0	}$&	228.1	$^{	+272.0	}_{	-435.1	}$&	300.0	$^{	+0.0	}_{	-280.0	}$&	32.5	$\pm$	14.6	&	2.2	&	189/198	\\
Ser~X--1	&	LMXB        	&	4.14E+21	&	36.118	&	4.842	&	11.1	$_{	-1.6	}^{+	1.6	}$	[4]	&	64.2	&	57.4	&	17.1	$^{	+0.6	}_{	-1.5	}$&	159.4	$^{	+340.6	}_{	-492.4	}$&	95.8	$^{	+204.2	}_{	-75.8	}$&	27.5	$\pm$	11.9	&	2.3	&	253/278	\\
XB~1832--330	&	LMXB        	&	9.24E+20	&	1.531	&	-11.371	&	9.6	$_{	-0.4	}^{+	0.4	}$	[8]	&	63.6	&	54.0	&	16.7	$^{	+1.0	}_{	-1.1	}$&	-212.6	$^{	+459.9	}_{	-287.5	}$&	102.1	$^{	+198.0	}_{	-82.1	}$&	25.7	$\pm$	10.5	&	2.5	&	151/181	\\
4U~0513--40	&	LMXB        	&	4.21E+20	&	244.510	&	-35.036	&	12.1	$_{	-0.3	}^{+	0.3	}$	[6]	&	23.7	&	28.7	&	\nodata	$^{		}_{		}$&	\nodata	$^{		}_{		}$&	\nodata	$^{		}_{		}$&	$<$ 56.1			&	\nodata	&	\nodata	\\
4U~0513--40$^{*}$	&	\nodata       	&	\nodata	&	\nodata	&	\nodata	&	\nodata	$_{		}^{		}$		&	\nodata	&	\nodata	&	16.5	$^{	+0.4	}_{	-0.5	}$&	-595.0	$^{	+340.1	}_{	-5.0	}$&	300.0	$^{	+0.0	}_{	-280.0	}$&	44.1	$\pm$	15.0	&	2.9	&	90/92	\\
LMC~X--4	&	HMXB        	&	1.17E+21	&	276.335	&	-32.529	&	49.97	$_{		}^{		}$	[12]	&	43.7	&	28.7	&	16.8	$^{	+0.6	}_{	-1.8	}$&	0.2	$^{	+383.7	}_{	-391.9	}$&	97.1	$^{	+202.9	}_{	-77.1	}$&	25.5	$\pm$	15.5	&	1.6	&	100/92	\\
4U~0919--54	&	LMXB        	&	6.16E+21	&	275.852	&	-3.845	&	5	$_{	-0.7	}^{+	0.8	}$	[18]	&	39.7	&	26.3	&	15.8	$^{	+1.6	}_{	-0.8	}$&	0.0	$^{	+500.0	}_{	-500.0	}$&	97.5	$^{	+202.5	}_{	-77.5	}$&	3.4	$\pm$	15.2	&	0.2	&	73/103	\\
MXB~1659--29	&	LMXB        	&	1.82E+21	&	353.827	&	7.266	&	11.5	$_{	-1.5	}^{+	1.5	}$	[6]	&	30.6	&	24.8	&	16.4	$^{	+1.2	}_{	-1.4	}$&	237.0	$^{	+263.0	}_{	-537	}$&	95.1	$^{	+204.9	}_{	-75.1	}$&	21.1	$\pm$	15.7	&	1.3	&	83/82	\\
4U~1746--371	&	LMXB        	&	2.63E+21	&	353.531	&	-5.005	&	11	$_{	-0.8	}^{+	0.9	}$	[22]	&	71.3	&	22.9	&	17.1	$^{	+0.4	}_{	-1.3	}$&	-272.7	$^{	+301.8	}_{	-227.3	}$&	141.0	$^{	+159.0	}_{	-120.9	}$&	38.3	$\pm$	17.5	&	2.2	&	93/94	
\enddata
\label{table:result}
\tablecomments{[1]\citet{jonker2004}
[2]\citet{trimble1973}
[3]\citet{sala2006}
[4]\citet{kuulkers2003}
[5]\citet{hynes2004}
[6]\citet{int-zand2001}
[7]\citet{caraveo2001}
[8]\citet{zurita2008}
[9]\citet{helton2008}
[10]\citet{ziokowski2005}
[11]\citet{mason1982}
[12]\citet{pietrzynski2013}
[13]\citet{galloway2006}
[14]\citet{savolainen2009}
[15]\citet{int-zand2003}
[16]\citet{nowak2008}
[17]\citet{kaplan2007}
[18]\citet{wang2004}
[19]\citet{rau2011}
[20]\citet{kong2000}
[21]\citet{hessels2007}
[22]\citet{paerels2001}
[23]\citet{muno2001}
[24]\citet{tetzlaff2011}
}
\end{deluxetable}
\clearpage
\end{landscape}

\clearpage

\begin{figure*}
\center
\includegraphics[width=\textwidth,height=0.7\textheight,angle=0]{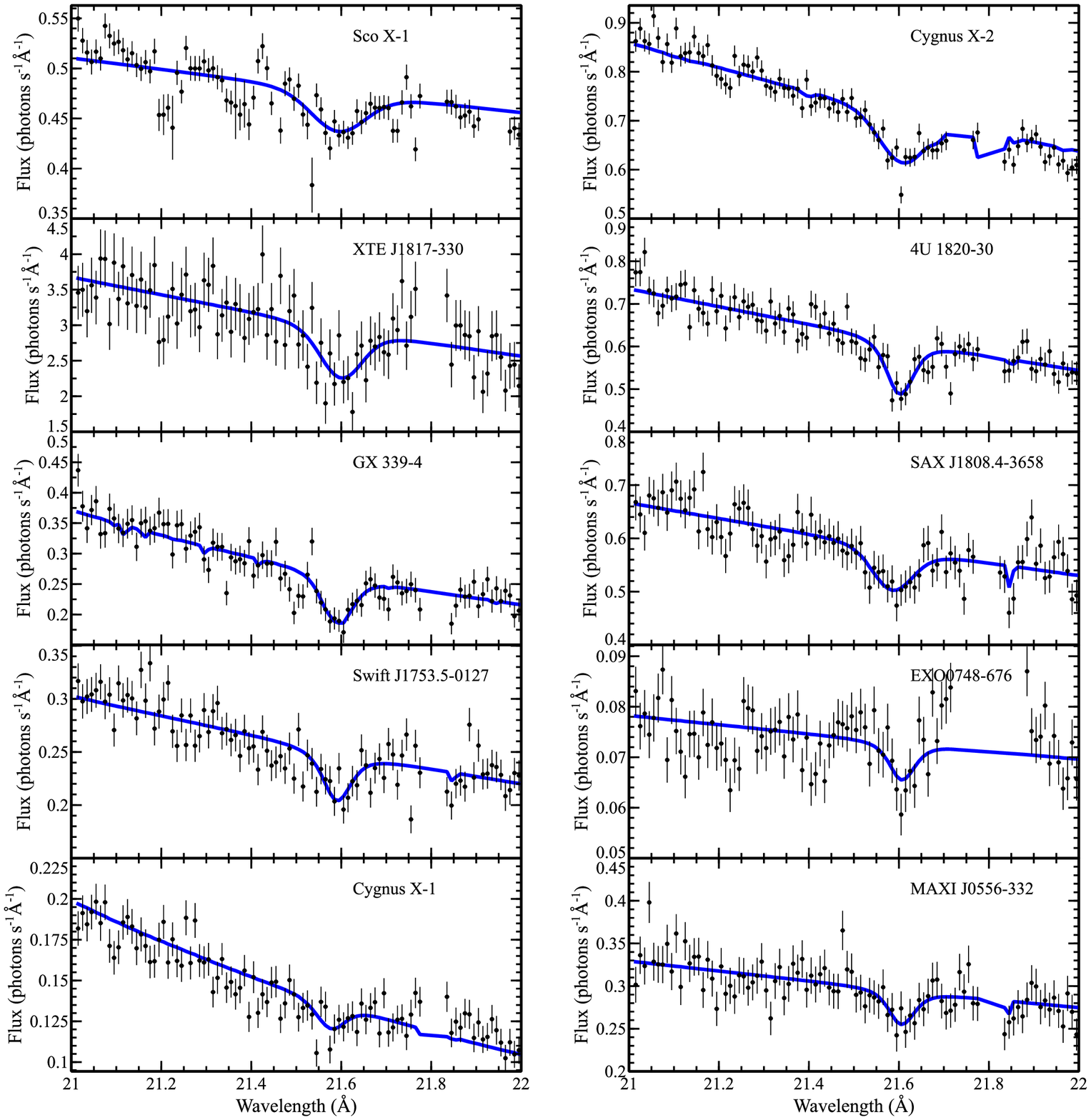}
\caption{RGS spectra between 21 and 22 \AA\ for each target. The blue line is the fitted continuum. This is only for demonstration purpose because for each target the spectra were jointly fitted with the same absorption line parameters but different continuum level. The structure at $\sim$ 21.8 \AA\ in each panel is an instrumental feature, and is ignored in the fit.}
\label{fig:spec1}
\end{figure*}

\begin{figure*}
\center
\includegraphics[width=\textwidth,height=0.7\textheight,angle=0]{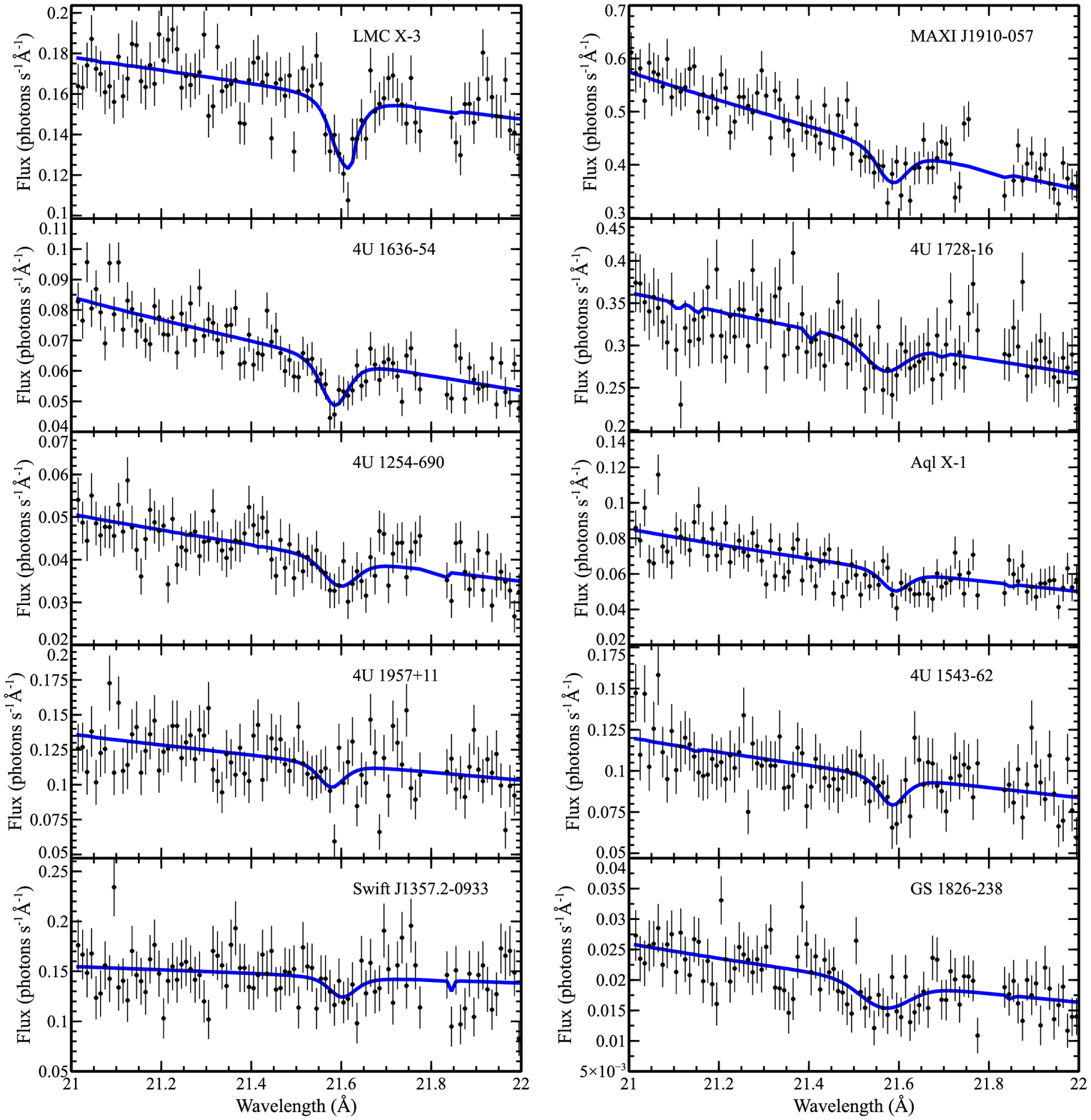}
\caption{Same as Figure \ref{fig:spec1} but for different targets.}
\label{fig:spec2}
\end{figure*}

\begin{figure*}
\center
\includegraphics[width=\textwidth,height=0.7\textheight,angle=0]{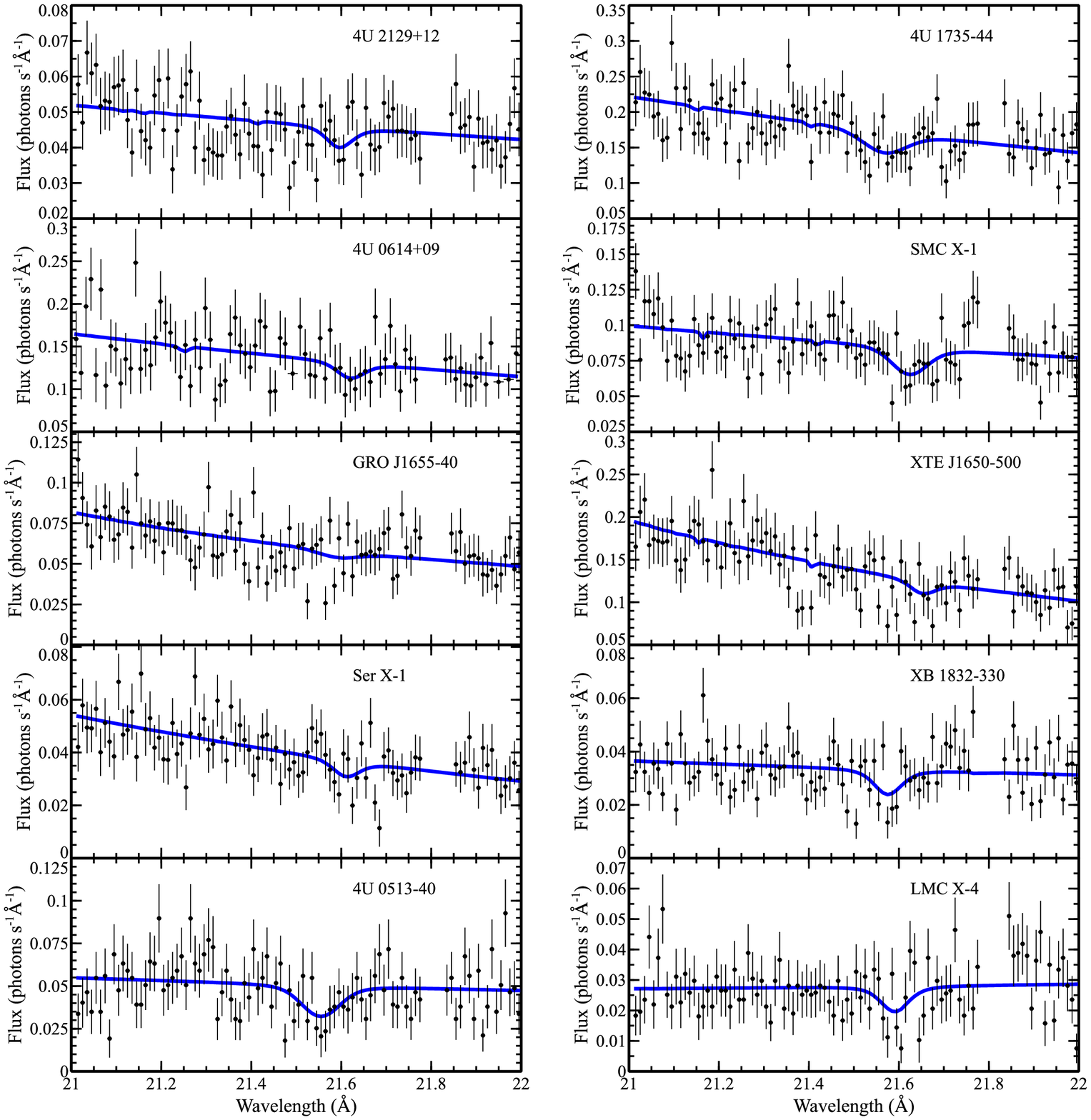}
\caption{Same as Figure \ref{fig:spec1} but for different targets.}
\label{fig:spec3}
\end{figure*}

\begin{figure*}
\center
\includegraphics[width=\textwidth,height=0.34\textheight,angle=0]{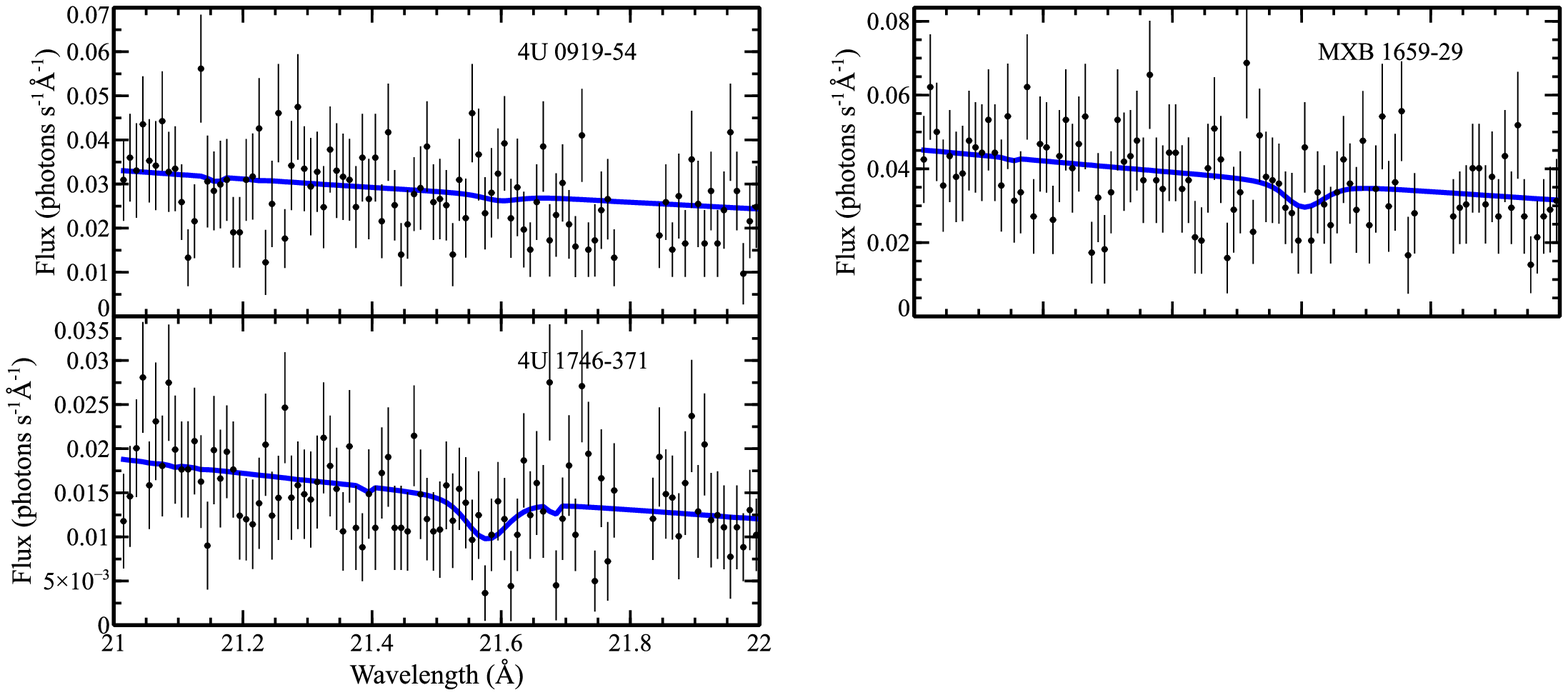}
\caption{Same as Figure \ref{fig:spec1} but for different targets.}
\label{fig:spec4}
\end{figure*}

\appendix
We present all the target spectra in Figure \ref{fig:spec1} - \ref{fig:spec4} in the 21--22 \AA\ wavelength range. The blue line in each panel is the model fitted continuum and absorption line. The model fitted continuum in these figures are only for display purpose because for each target the continuum changes among different observations. There is a detector feature at about 21.82 \AA\, so we have excluded this region in fitting the spectrum.

\end{document}